# Is the Photon Really a Particle?

**Paul A. Klevgard, Ph.D.**

pklevgard@gmail.com

**Sandia National Laboratory, Ret.**

## Abstract

Photons deliver their energy and momentum to a point on a material target. It is commonplace to attribute this to particle impact. But since the in-flight photon also has a wave nature, we are stuck with the paradox of wave-particle duality. It is argued here that the photon's wave nature is indisputable, but its particle nature is open to question.

Photons deliver energy. The problem with invoking impact as a means of delivery is that energy becomes a payload which in turn requires a particle. This assumes that energy is always a payload and there is but one mode of energy delivery; surely two unsupported assumptions.

It should be possible to explain photon termination without invoking particle impact. One approach offered here is to question the assumption that the photon is a unitary object. Perhaps the photon has two linked-but-distinct identities: one supporting wave behavior and the other supporting discrete behavior. It is the latter that might imitate particle impact.

Keywords:      wave-particle duality; photon; dualism; radiation, kinetic energy

***************************

Sole author; no conflicts of interest.

This research did not receive any specific grant from funding agencies in the public, commercial, or not-for-profit sectors.

**1.0     Introduction**

How can light be both a particle and a wave? This question has been with us for over a century and it is hard to detect any progress toward a resolution. Most theoretical physicists and philosophers of physics have moved on to other, more career-rewarding issues; wave-particle duality has become "baked-in" to their world view. Modern optics and photonics have spurred many duality experiments with light, but this research rarely ventures into theoretical issues. We are left trying to adapt the usual interpretations of quantum particle physics to radiation.

The Copenhagen interpretation does not attempt to explain duality for the electron or for the photon; rather it insists that duality is fundamental and beyond our ken. It is the experimenter's choice of measurement devices that determines whether waves or particles are detected. For Bohr the description of any quantum phenomenon always requires both the particle picture and the wave picture [1, p. 90]. It has been said that Complementarity solves the dualism dilemma "…by simply accepting both of the dilemma's horns [2, p.123]."

The de Broglie-Bohm pilot wave theory was devised to explain how rest mass particles (e.g., electrons) are guided by the wave function. The theory has been applied to the photon [3], [4], [5], [6], but various issues arise. For example, Schrödinger's wave equation applied to radiation devolves into Maxwell's equation. The whole ontology of Bohm's theory is more focused on particle and wave rather than on mass and energy. Hence if the electron and the photon both behave like waves, then their differences (mass vs. massless) may be ignored. In addition, it is not clear how the photon can have that "complex and subtle inner structure" [7, p.37] which Bohm attributes to mass-bearing particles.

Quantum Field Theory (QFT) says that all mass-bearing particles and the photon are excitations of some underlying, quantized field. Supporters [8], [9], [10] of this view claim it resolves wave-particle dualism. But these quantized fields are operator valued and "…it is exceptionally unclear which parts of the formalism should be taken to represent anything physical in the first place [11]." Aside from this problem, interpreting oscillatory field modes as particles is to keep both discrete and continuous features at the ready, making them available for invocation when convenient. It is ontologically problematic to encompass the photon (no rest mass, no trajectory) and the electron (rest mass and trajectory) with one model. Despite its many successes, QFT is not well adapted to explain radiation dualism.

## 2.0     Photon as Wave

Electrons and photons both exhibit dualism when traversing a double slit. This prompts most to assume there is a common explanation for both. Perhaps there is, but that is a larger question. Our purpose here is more modest and there is some value in treating the two cases separately. It is not clear how, or if, the wave function and Heisenberg uncertainty apply to the in-flight photon, whereas they certainly apply to the electron.

So our dualism focus here is just on the photon. What evidence tells us the photon is a wave and what tells us it is a particle?

The wave (continuous) nature of radiation seems indisputable. Since EM radiation is oscillatory, its progression in space yields the waveform. Radiation exhibits interference, diffraction, rarefaction and superposition. Radiation waves are on secure grounds mathematically, both via wave (physical) optics and Maxwell's equations. Wave diffraction patterns of coherent light can be modeled mathematically from relatively few parameters.

The particle nature of radiation is not mathematically grounded, at least nothing comparable to Maxwell's equations. Radiation is quantized and delivered (to matter) in discrete energy chunks, but that does not prove that the chunk received is a particle. The particle concept assumes that energy can only be a payload, i.e., something carried by something else. The identification of the photon as a particle seems to rest upon its delivery of energy and momentum to a point in space and time. The analogy with particle impact is obvious, perhaps too obvious. It was certainly decisive for Richard Feynman who cites the individual arrival of photons on a detector and concludes: "[w]e know that light is made of particles… [12, p.14]."

P.A. Klevgard, May 2021

But this reasoning-by-analogy is a classic instance of underdetermination.[1] Working physicists are a pragmatic lot. They write equations for the delivery of energy and force to material particles. Abraham Pais [14, p. 350-1] writes that although the photon has zero mass, physicists "… nevertheless call a photon a particle because, just like massive particles, it obeys the laws of conservation of energy and momentum in collisions, with an electron say (Compton effect)."

So the wave nature of the photon is convincing, whereas its particle nature is open to question. Presently an alternate explanation of radiation's "discrete arrival" will be made. But photon arrival is probabilistic; we first we need to reconcile the arrival of photon energy with photon probability.

### 3.0   Probability Wave as a Photon Identity

The function of radiation is the transmission of kinetic energy over space. Radiation quanta, photons, will only terminate on a material object and they do so discretely. The energy of the photon's origin is delivered to its termination point undiminished. But in-flight photons are subject to diffraction, yielding multiple space paths with the reception of a photon on a particular path being probabilistic.

We are all familiar with a double slit diffraction pattern built over time by photons arriving separately. Individually the photons are entirely random, but in aggregate they reveal that photons conform probabilistically to a waveform. So coherent photons (frequency and phase [crests] aligned) all share the same waveform which governs their aggregate behavior, but not their individual termination. The nature of this probability wave is unclear, but its (diffractable) presence in space – which we can model mathematically – indicates it is physically real.

So one photon identity is a diffracting, rarefying probability wave travelling over space. Its other identity is a quantized energy which never diffracts or rarefies.

---

[1] "For any theory, T, and any given body of evidence supporting T, there is at least one rival (i.e., contrary) to T that is as well supported as T [13, p.271]."

### 4.0     Energy as the Second Photon Identity

Work done on a charge creates photon energy. Two waveforms are also created: 1) diffractable probability waves; and 2) electric-magnetic oscillations that are a consequence of charge acceleration. Both of these waves expand, diffract and rarefy indefinitely in space, but photon energy neither rarefies nor diffracts there. In addition, at photon termination these waves must collapse nonlocally in space and this would be impossible were they carriers of energy.

From E = *hf* we know that photon energy is oscillatory. The photon's two space waves (probability and EM) oscillate, but if neither of them carries energy, then what else is oscillating and where does it reside?

Perhaps there is such a thing as pure oscillation for photon energy. While this is speculative, physics already has vacuum state fluctuation as the oscillation of "nothing" or the oscillation from "nothing." Without invoking the overhead of QFT (undetectable fields and the uncertainty principle), let's suppose that photon pure oscillation is the consequence of work done on a charge. Work done is kinetic energy and the latter might then simply occur in its own right as oscillation; this makes occurrence/energy the ontological equal of existence/mass which has some support from $E = mc^2$.

Oscillation involves cycles and they require time, so pure oscillation will occupy (require) time. And to occupy an interval in a dimension is to reside there. So oscillation energy may reside in time, but it will still be available for space interaction. Quantized mass/energy objects in general may occupy one dimension and still be accessible for the other dimension; your desk occupies space but is still present for interaction over time. Hence photon oscillation energy in time would still be available for interaction over space.

### 5.0     Two Identities

So perhaps the photon has two identities. Each identity addresses a specific feature of the photon that has long puzzled commentators.

First, the photon has no space trajectory; it famously "follows all paths" like a wave. Acknowledging that one photon identity is a probability wave appears to account for this singular feature. So while the photon is not actually on every path, it does have the probability of terminating its energy on any available path. Strictly speaking, it is photon probability that "follows all paths." Photon energy reception on a target is then governed in aggregate by the photon's waveform, space-progressing, probability identity.

A second puzzling feature is the non-diffraction of photon kinetic energy by space objects (slits, pinholes). Granting photon energy a time-residing, oscillatory nature could resolve this issue. Photon energy cycles residing in time cannot be divided in space, just as a material object residing in space cannot be divided in time.[2] Residing in time renders photon energy immune from space diffraction. What diffracts in space are the photon probability waves.

So it is possible there are two photon identities that are functionally different: space-progressing probability waves and time-residing energy, with the latter still accessible for space. Photons can only terminate on matter; it could be that this involves the transfer of time-residing photon oscillation energy to space-residing rest mass. And the only way the two can meet/intersect is via an event which combines kinetic energy from time with rest mass from space. Ontologically, this is the intersection of occurrence (energy) with existence (mass) mediated by probability.

<u>Such an event/intersection is discrete in both space and time since the two actors are orthogonal in terms of where they reside</u>. The transfer of time-residing kinetic energy to space-residing matter is necessarily localized, event based and discontinuous. And it mimics particle impact.

## 6.0    Particle Impact vs. Energy-Matter Intersection

Photon reception delivers both energy and momentum discretely to the target rest mass, just as an impacting particle does. But there is a big difference between the two cases. A particle impacting on a material target is an event between two mass-based objects; each has a trajectory relative to the other's frame of reference, whereas a photon has neither a trajectory nor a frame of reference. Photon energy impinging on a material target is an event (absorption) created by the intersection of ontological opposites, one mass-based, the other not. We don't invoke the particle concept for photon emission; we should not invoke it for its inverse, photon absorption. Arguing that the photon is a special kind of particle exempt from normal rest mass features is unphysical.

---

[2] Material objects progress (persist, age) in time. They do not occupy (reside in) time. You can't divide anything (your rest mass desk or photon energy) in a dimension where it does not reside.

**7.0     Conclusions**

The discrete nature of photon termination can be explained without recourse to the simplistic analogy of particle impact. Alternative explanations are possible; explanations that don't treat photon energy as the payload of a massless particle.

The possibility presented here involves a new way of regarding photon energy plus the presumption of photon identities. Entities in general possess two identities. A material object/entity has its rest mass as one identity and its stored energy (thermal or intrinsic) as a second identity. Granting the photon two identities allows it to function differently in space and in time. If photon oscillatory energy resides in time, its reception on a space target that is dimensionally orthogonal must be discrete.

As outlined, the wave nature of the photon has been well-established by experiment and theory. This is not true for the concept of the photon as particle. Radiation occurs; matter exists. Regarding the photon as a particle is to apply the ontology of existing matter to occurring radiation.

P.A. Klevgard, May 2021

# Appendix A

If Nature embraces symmetry, then (photon) energy occurring in time would have its counterpart in mass existing in space. But it seems unlikely that our material instruments can confirm that photon oscillation energy resides in the time dimension. Granting credence to this idea depends upon how well it explains other phenomena. As Einstein argued, a correct theory is one that addresses many problems: gestaltic verification. Placing photon energy in time resolves photon dualism by eliminating impact as an explanation for reception. Other photon puzzles also benefit.

**Entanglement:**

A high energy photon entering a crystal may split into two lesser but entangled photons. Assume these daughter photons have their energies bonded in the time dimension while their separate probability-of-reception waves fan out in space. If the waves of one photon initiate reception on a detector, then that photon's spin is determined, up or down. This simultaneously determines the (opposite) spin of its entangled, time-residing partner. Nonlocal space communication does not take place since spin coordination occurs in time.

Electrons may also have their time-residing energies entangled; such entanglement may even survive some electron space separation.

**Constant velocity of light**:

With photon energy in time, what progresses in space are two immaterial waves: EM waves and probability-of-reception waves. The velocity of these waves is their wavelength times their frequency. If you move toward a light source, you diminish the wavelength but increase the frequency. If you move away from the light source, you increase the wavelength but diminish the frequency. In both cases the wave (phase) velocity stays the same.

**Details**:

doi.org/10.1016/j.ijleo.2021.168180; also https://arxiv.org/abs/2108.08222